\documentclass[12pt,english]{article}

\newcommand{\CL}{{\cal L}}

\makeatletter
\newcommand*{\rom}[1]{\expandafter\@slowromancap\romannumeral #1@}
\makeatother

\usepackage[latin9]{inputenc}
\usepackage{geometry}
\usepackage{verbatim}
\usepackage{prettyref}
\usepackage{amsmath}
\usepackage{amssymb}
\usepackage{enumitem}
\usepackage{cases}
\usepackage{graphicx}
\usepackage{caption}
\usepackage{subcaption}
\usepackage{tikz}
\usepackage{tikzsymbols}
\usepackage{physics}

 \usepackage{slashed}
\usepackage{esvect}
\usepackage{dsfont}

\usepackage{color}
\definecolor{darkgreen}{rgb}{0,0.5,0}
\definecolor{darkblue}{rgb}{0,0,0.6}
\definecolor{npurple}{rgb}{0.4,.2,0.7}
\RequirePackage[colorlinks=true, urlcolor=black, menucolor=black, linkcolor=purple, citecolor=purple]{hyperref}
 	
\numberwithin{equation}{section}
\numberwithin{figure}{section}
\numberwithin{table}{section}

\def\CZ{{\cal Z}}

\newcommand{\SO}{\mathop{\mathrm SO}}

\newcommand{\be}{\begin{equation}}
\newcommand{\ee}{\end{equation}}
\let\vec\mathbf

\addtocounter{section}{0}

\title{\textbf{Low-temperature thermal corrections\\ to a superfluid's equation of state}}
\author{\href{mailto:ik2436@columbia.edu}{Ioanna Kourkoulou}, \href{mailto:a.nicolis@columbia.edu}{Alberto Nicolis}, and 
\href{mailto:k.parmentier@columbia.edu}{Klaas Parmentier}\\\it \small Center for Theoretical Physics, Department of Physics,\\ \it\small
Columbia University, New York, NY 10027, USA}
\date{}
\begin{document}
\fontseries{mx}\selectfont	
\maketitle
\begin{abstract}
Using effective field theory methods, we calculate the low-temperature phonon thermal partition function for a generic superfluid, thus providing the leading thermal corrections to a superfluid's equation of state, thermodynamic quantities, and conserved currents. We confirm the correctness of the classic two-fluid model by matching to the effective field theory for that model. Where there is an overlap, our results agree with those of Carter and Langlois, which were derived by different methods. We also discuss generalizations and special cases, including the non-relativistic limit and the conformal superfluid. 
\end{abstract}
\thispagestyle{empty}
\newpage
{\hypersetup{linkcolor=black}
\tableofcontents}
\thispagestyle{empty}
\addtocounter{page}{-2}

\newpage
\section{Introduction}
A relativistic superfluid at zero temperature can be described by an EFT Lagrangian for a single Poincar\'e scalar $\psi$ \cite{Son:2002zn},
\begin{equation}\label{eq:superfluid}
    \CL_0 = P (X) \; , \quad X = -\partial_\mu \psi \partial^\mu \psi,
\end{equation}
where $P$ is a function related to the equation of state. Namely, it gives the pressure as a function of the chemical potential.

Expanding the field in fluctuations around a given background configuration $\bar \psi(x)$, $\psi = \bar \psi + \pi$, one finds the  dynamics for the phonons $\pi$. In particular, at the quadratic level, one obtains
\begin{equation}\label{eq:quadpi}
    \CL_\pi  = \frac12 Z^{\mu\nu}\partial_\mu \pi \partial_\nu \pi, \qquad Z_{\mu\nu} = 4P''( \bar{X} ) \,  \partial_\mu \bar \psi \, \partial_\nu \bar \psi - 2 P'( \bar{X}  ) \, \eta_{\mu\nu},
\end{equation}
where $\bar{X}$ is the scalar quantity $X$ as in \eqref{eq:superfluid}, evaluated on the background value $\bar \psi$ of $\psi$.  In what follows we will simply denote it by $X$ as well, to avoid clutter, since both quantities are the same at the order of interest.

At finite, but low, temperatures $T$, the system will consist of both the superfluid background and a thermal gas of phonons, which behaves as a regular fluid. This is the celebrated two-fluid model. To compute the thermodynamic properties of this gas, and to find the associated EFT description of the two-fluid system, one starts by calculating the thermal partition function for the phonon gas. At low $T$, knowledge of the quadratic Lagrangian \eqref{eq:quadpi} suffices for this purpose. The reason is that phonons, owing to their Goldstone boson nature, are derivatively coupled. Their interactions are thus weaker and weaker at lower and lower energies. This implies that, to lowest order in $T$, phonons can be treated as free particles.

In cases where $\partial_\mu \bar \psi$ is timelike, one can go to the superfluid rest frame. There, the phonons move at the speed of sound $c_s$, given by
\begin{equation}\label{eq:sound}
    c_s^2 = \frac{P'(X)}{P'(X)+2P''(X) X}, \qquad \frac{dc_s^2}{dX} = \frac{-2P'P'' + 2 P'' {}^2 X - 2P' P''' X}{(P' + 2P'' X)^2},
\end{equation}
where we included the first derivative for future reference. Starting from this, Carter and Langlois \cite{Carter:1995if} then derived the partition function for a thermal gas of phonons and the associated equation of state for a cold relativistic superfluid. 
To do so, they used traditional thermodynamic and hydrodynamic methods. Here, we take a field theoretical approach, and, by doing so, arrive at the result in a more modern language. This allows us in turn to apply the results in a more general setup; it makes, for instance, the generalization to arbitrary dimensions straightforward.

In EFT, the regular fluid component is described by an $\SO(3)$ triplet of scalars $\phi^I$ \cite{Nicolis:2015sra}. A superfluid-fluid mixture should then be described in terms of the building blocks 
\begin{equation} \label{sf blocks}
    X = -\partial_\mu \psi \partial^\mu \psi, \quad 
    y= \frac{1}{b}\epsilon^{\mu\nu\rho\sigma}\partial_\mu\psi \partial_\nu \phi^1 \partial_\rho \phi^2 \partial_\sigma \phi^3,\quad b= \sqrt{\det (\partial_\mu \phi^I\partial^\mu \phi^J)} \; ,
\end{equation}
through a general Lagrangian
\be \label{f}
{\cal L} = F (X, y, b) \; . 
\ee
The Carter-Langlois equation of state interpreted in EFT language is indeed of precisely this form. Concretely, for low $T$, the correction to the superfluid EFT takes the form \cite{Nicolis:2011cs}
\begin{equation}\label{eq:finT}
    \CL = \CL_0 + \mathcal{L}_1 = P(X) - 3  \bigg[ \frac{b^4 }{ c_s(X) } \Big(1- \big(1-c^2_s(X) \big) \frac{y^2}{X} \Big)^{2} \bigg]^{1/3}.
\end{equation}

Instead of the thermodynamic treatment of \cite{Carter:1995if}, we aim to find an independent and more direct field theory justification for \eqref{eq:finT}. On the one hand, this sheds light on its physical content, and does so in a perhaps more digestible and modern EFT notation. On the other hand, it allows for increased generality. In particular, we can consider superfluid backgrounds for which a standard interpretation would correspond to a superluminal velocity. In this case the Carter-Langlois strategy of going to the superfluid's rest frame does not apply. However, these setups involving spacelike $\partial_\mu \bar \psi$ turn out to be well-defined, causal, and stable - simply, the usual interpretation in terms of superfluid velocity fails in that case. We explore such setups in a companion paper \cite{kourkoulou}, which in fact motivated some of the work presented here. Generalizing the low-$T$ Lagrangian to spacelike superfluid backgrounds was necessary for studying the stability properties and subtleties of such setups; the derivation we show below provided us with the required tools, which were previously unavailable.

The idea is that the general superfluid EFT \eqref{eq:superfluid} already contains all the information we need.  From it, we  calculate the phonon thermal partition function,
which, by differentiation,  yields the finite-temperature corrections to the  energy-momentum tensor and to the $U(1)$ current.
On the other hand, a two-fluid  EFT like \eqref{f} predicts a specific energy-momentum tensor and $U(1)$ current for the phonon gas, in terms of $F$ and its derivatives. By matching these physical observables we  determine the function $F$, confirming the result in \eqref{eq:finT}, but also extending it to spacelike superfluid-like backgrounds, and to different dimensions.

We also consider a number of different limits and special cases of our results.
As a byproduct, by studying the non-relativistic limit we also confirm the results of \cite{Nicolis:2017eqo, Esposito:2018sdc} regarding the mass carried by sound waves and phonons.

%
%

Similar arguments appeared in \cite{Alford:2012vn} for superfluids with a weakly coupled (complex) $\phi^4$ UV completion. However, the advantage of starting from the low-energy EFT \eqref{eq:superfluid} for a low-temperature computation is two-fold: on the one hand, the computation is simpler because at low temperature only the low-energy degrees of freedom get excited, and so it is convenient to have a starting point that only involves those; on the other hand, the results one finds are more general, since they only follow from the symmetries and the spontaneous breaking thereof, and are thus independent of any specific microphysics model.

\vspace{.5cm}
\noindent
{\it Note added\,}:  A very recent paper, ref.~\cite{Gouteraux:2022kpo}, has results that partially overlap with some of ours, specifically with those of sect.~\ref{Cherenkov}. 

\vspace{.5cm}
\noindent
{\it Notation and conventions\,}: Throughout the paper, we work in natural units, $\hbar = c = k_B =1$, and with the mostly-plus signature for the spacetime metric $\eta_{\mu\nu}$.  Our initial derivation is in 3+1 dimensions, but later we generalize to generic $D=d+1$. We call our system a {\it relativistic} superfluid, simply because we take into account special relativity.  Our results apply equally well to standard `non-relativistic' superfluids, such as liquid helium-4 (see e.g.~\cite{Nicolis:2015sra}).  We will use the standard notation $\mu$ for chemical potential, which should not be confused with the spacetime index appearing for instance in the metric $\eta_{\mu\nu}$. This will be clear from the context.

\section{The partition function}\label{partition}

Our starting point will be the Lagrangian \eqref{eq:quadpi}, in the presence of a constant background 
\be
\partial_\alpha \bar \psi(x) \equiv V_\alpha = {\rm const} \; .
\ee 
For timelike $V_\alpha$, this is the generalization to a generic frame of having a constant chemical potential $ \mu$, which in the rest frame of the superfluid corresponds to $V_\alpha =  \mu  \, \delta_\alpha ^0$. For spacelike $V_\alpha$, it is an altogether different system: the superfluid interpretation fails in that case, since a spacelike $V_\alpha$ would imply a spacelike superfluid four-velocity field. We leave it to our companion paper \cite{kourkoulou} to discuss  the physical relevance of the spacelike case. For the purposes of this paper, we can perform our field-theory computations without committing to a specific $V_\alpha$.

Clearly, not all possible $Z^{\mu\nu}$ in \eqref{eq:quadpi} correspond to a stable system. It can be shown  that (perturbative) stability is equivalent to the conditions \cite{NR}
\be \label{stability}
Z^{00} > 0 \, \quad \mbox{and}  \quad  G^{ij} \equiv Z^{0i}Z^{0j}- Z^{00}Z^{ij} \succ 0 \; .
\ee
So, from now on we will assume that $V_\mu$ is such that these conditions are obeyed. 
In fact, it is sometimes useful to consider a {\em stronger} condition for stability \cite{NR},
\be \label{stronger stability}
Z^{00} > 0 \, \quad \mbox{and}  \quad  (-Z^{ij}) \succ 0 \; ,
\ee
which says that kinetic energy and gradient energy should both be positive, regardless of the size of the $\dot \pi \, \vec  \partial_i \pi$ mixing terms. The difference between the two criteria is discussed at length in \cite{Dubovsky:2005xd}, and it amounts to the system admitting two consistent but inequivalent quantizations. This latter, stronger condition recognizes that Cherenkov emission can be thought of as a frame-dependent instability. We discuss this and the implications for our results in sect.~\ref{Cherenkov}.

We now want to compute the phonon partition function ${\cal Z}(T, V_\alpha)$ at finite temperature $T=1/\beta$ and for generic $V_\alpha$. As for any bosonic free field theory, its log per unit volume
\footnote{With an abuse of notation, we will refer to this particular quantity as ``the partition function". In fact, it is minus the free energy density divided by the temperature.} takes the standard integral form 
\be
z(T, V_\alpha) \equiv \frac{\log \CZ (T, V_\alpha)}{\cal V} = - \int \frac{d^3k}{(2\pi)^3} \log(1-e^{-\beta \omega_{\vec k}}) \; ,
\ee
where ${\cal V}$ is the standard $3$-volume factor, and $\omega_{\vec k}$ is the frequency of a phonon of momentum ${\vec k}$.
Rewriting the equations of motion for the $\pi$ excitations in Fourier space, we find the  dispersion relation 
\begin{equation}
    Z^{00}\omega^2 - 2 Z^{0i}\omega k_i + Z^{ij}k_i k_j = 0 \; .
\end{equation}
%
The positive frequency solutions take the form 
\begin{equation}\label{eq:disp}
    \omega_{\vec k}  = \frac{1}{Z^{00}}\big( Z^{0i} k_i + \sqrt{G^{ij} {k}_i {k}_j} \, \big) \; ,
\end{equation}
where $G^{ij}$ is defined in \eqref{stability}.


Given that $G^{ij}$ is symmetric and positive definite, its (positive definite) square root is well defined. We thus write
\be
p^i \equiv (G^{1/2})^{ij} \, k_j \; , \qquad C_i \equiv (G^{-1/2})_{ij} \, Z^{0j} \; .
\ee
This leads to the partition function
\begin{equation}\begin{split}\label{eq:logZderiv}
    z (T,  V_\alpha)
    &= -   \frac{1}{\sqrt{\det G}} \int \frac{d^3p}{(2\pi)^3}  \log(1- e^{-\frac{\beta}{Z^{00}}(\vec C \cdot \vec p + |\vec p| \, )})\\
    & = \frac{\pi^2}{90 \beta^3} \frac{(Z^{00})^3}{\sqrt{\det G} \big(1- |\vec C|^2 \big)^2} \\
    & = \frac{\pi^2}{90 \beta^3} \frac{(Z^{00})^3}{\sqrt{\det G} \big(1- (G^{-1})_{ij} Z^{0i} Z^{0j} \big)^2} \;  .
\end{split}
\end{equation}
In the next section we shall need to take derivatives of $z$ with respect to $Z^{\mu\nu}$. It is therefore useful to keep the latter explicit. Using the definition of $G^{ij}$, after some straightforward but tedious matrix gymnastics, the above can be rewritten as
\begin{equation}\label{eq:logZ}
    z(T, V_\alpha)=   \frac{\pi^2 T^3}{90} \frac{\sqrt{-\det A}}{ \big( A_{00} \big)^2 } \; , \qquad  A_{\mu\nu} \equiv (Z^{-1})_{\mu\nu} \; , 
\end{equation}
where we introduced the matrix $A$, the inverse of $Z$, for future convenience.
This is the thermal partition function for phonons described by the general quadratic Lagrangian \eqref{eq:quadpi}. 

We'd like to make this more covariant looking by introducing the four velocity of the lab frame. So far we have been doing thermodynamics in a specific frame, that where we introduced the temperature $T$. So, all zero indices should be interpreted as standing for the time direction in that particular frame. Since the four velocity of that frame, in that same frame, is $u^\mu = (1,\vec 0)$, we can convert all lower zero indices to lower Lorentz indices contracted with $u^\mu$. The resulting expression,
\begin{equation}\label{z covariant}
    z(T, V_\alpha)=   \frac{\pi^2 T^3}{90} \frac{\sqrt{-\det A}}{ \big( A_{\mu\nu} \, u^\mu u^\nu \big)^2 } \; ,
 \end{equation}
 is then valid in all frames. 

It is also useful to express this directly in terms of the background vector field $V_\mu = \partial_\mu \bar \psi$ and the sound speed \eqref{eq:sound}. Rewriting $Z^{\mu\nu}$ in  \eqref{eq:quadpi} as
\be \label{Zmn}
Z^{\mu\nu} = -\frac{2 P'}{c_s^2} \bigg[\big(1- c^2_s \big) \frac{V^\mu V^\nu}{V^2} + c^2_s \, \eta^{\mu\nu} \bigg]  \; ,
\ee
where $P'$ and $c_s^2$ are both evaluated at $X = -V^2$, 
we find
\be \label{A}
A_{\mu\nu} = \frac{1}{2 P'}\bigg[ \big(1- c^2_s \big)   \frac{V_\mu V_\nu}{V^2} - \eta_{\mu\nu}  \bigg] \; ,
\qquad 
\det A = - \frac{c_s^2}{(2 P')^4} \; ,
\ee
so that our partition function can be rewritten simply as
\be
z(T, V_\alpha)=   \frac{\pi^2 T^3}{90} \frac{c_s \, X^2}{ \big[X-(1-c_s^2) y^2\big]^2} \; ,
\ee
where $X$ and $y$ are defined as
\be
X=-V^2\; , \qquad y = u^\mu V_\mu \; .
\ee
We stress again that $c_s$ is implicitly a function of $X$, see eq.~\eqref{eq:sound}.

As a nontrivial check, notice that for a purely timelike $V_\alpha$, which is the standard case of a chemical potential defined in the same frame as that associated with the temperature, this reduces to the well known phonon partition function
\be
\frac{\pi^2 \, T^3}{90 \, c_s^3} \;  .
\ee

Finally, computing the entropy density, $s$, will come in handy. The standard relationship with the partition function yields, for our case, 
\be \label{s}
s = \frac{\partial \big (T z \big)}{\partial T} = 4 z (T, V_\alpha) \; .
\ee

\section{Thermal averages of conserved currents}
We now compute some physical quantities from the phonon partition function. We are particularly interested in the expectation values of the $U(1)$ current $J^\mu$ and of the stress-energy tensor $T^{\mu\nu}$, for two reasons. 

The first reason is that their expressions in terms of our background quantities $T$ and $V_\mu$ provide the most complete and unambiguous way to express the equation of state of the system: for a superfluid at finite temperature, there are in general two independent flows, and thus all standard quantities like the energy density, the number density, the pressure, etc., will depend on which one of these two flows is taken as defining ``the" rest frame of the system. By displaying instead the full covariant expressions for $J^\mu$ and $T^{\mu\nu}$, one completely bypasses this ambiguity. 

The second reason is that, by applying  Noether's theorem directly to the finite-temperature superfluid effective theory, one can also compute $J^\mu$ and $T^{\mu\nu}$ in terms of the EFT action \cite{Nicolis:2011cs}. By demanding that these match the expressions we now derive from the partition function, we will be able to derive the finite-temperature effective action, thus confirming the results of \cite{Nicolis:2011cs}, and extending them to the case of spacelike $V_\mu$. As we will see, the matching will provide quite a nontrivial check of the structure of the effective theory \eqref{f}. 

\subsection{The fundamental building blocks}

As we will see shortly, within the free-phonon approximation we have been using, our expectation values are  just suitable linear combinations of the entries of
\be
\Pi_{\mu\nu}  \equiv \langle \partial_\mu  \pi \, \partial_\nu \pi \rangle 
\; ,
\ee
where the two fields are evaluated at the same spacetime point.
So, let's start by computing this. We can do so directly from the partition function that we found above.

Standard functional methods imply
\footnote{It is enough to think of the partition function as the Euclidean path integral
\be
e^{{\cal V} z} = {\cal Z} = \int_\beta D\pi (x_E)\,  e^{-  \int d^4 x_E \, \frac12  Z_E^{\mu\nu} \partial^E_\mu \pi \partial^E_\nu \pi} \; ,
\ee
where Euclidean time $t_E = i t$ has period $\beta = 1/T$, and $Z_E^{\mu\nu}$ is the Euclidean version of $Z^{\mu\nu}$, i.e.~$Z_E^{00}=Z^{00}$, $Z_E^{0j} = -i Z^{0j}$, and $Z_E^{ij} = - Z^{ij}$. Deriving with respect to $Z^{\mu\nu}$ and keeping track of signs and $i$'s yields directly eq.~\eqref{Pi}.
}
\be \label{Pi}
\Pi_{\mu\nu} = 2T \frac{\partial z}{\partial Z^{\mu\nu}} = -2T \, A_{\mu\alpha} A_{\nu\beta} \, \frac{\partial z}{\partial A_{\alpha\beta}} \; .
\ee

Differentiating our expression \eqref{z covariant}, we get
\begin{align} 
\Pi_{\mu\nu} & =   - 2T \cdot z \cdot \bigg[ \frac{1}{2} A_{\mu\nu}- 2 u^\alpha u^\beta A_{\alpha\mu} A_{\beta\nu}/ A_{00}\bigg] \\
&  = - \frac{\pi^2 T^4}{90}  \cdot \sqrt{-\det A} \cdot \frac{\big(A_{\mu\nu} A_{\alpha\beta} - 4 A_{\alpha\mu} A_{\beta\nu}\big) u^\alpha u^\beta}{ \big( A_{\gamma \delta}\, u^\gamma u^\delta \big)^3 } \; \label{Pi result} .
\end{align}


Using \eqref{A}, we get
\begin{align} 
\Pi_{\mu\nu}  = \alpha \, 
&  \Big\{  -(1-c_s^2) \big[X + 3(1-c_s^2) y^2  \big]  \cdot V_\mu V_\nu \\
& -  4 X^2  \cdot u_\mu u_\nu \\
&-4 (1-c_s^2) X y  \cdot (u_\mu V_\nu+V_\mu u_\nu) \\
& + \big[(1-c_s^2)y^2 -X \big] X   \cdot \eta_{\mu\nu}\Big\} \; ,
\end{align}
where we isolated the four possible independent tensor structures, 
and the overall coefficient is a given function of $X = -V^2$ and of $y = u^\mu V_\mu$,
\be
\alpha \equiv  - \frac{\pi^2 T^4}{90}  \frac{c_s X}{(2P')\big[X- (1-c_s^2)y^2 \big]^3} 
\ee

Things are getting complicated, so, as a check that we have not made mistakes so far, we can consider the simplest possible case---that of a superfluid at rest in the lab frame defined by $u^\mu = (1,\vec 0)$, that is, $V_\mu \propto u_\mu$. 
After some major cancellations, we get
\be 
\Pi_{00} = \frac{\pi^2 T^4}{30} \frac{1}{(2P') c_s}  \; , \qquad \Pi_{0i} = 0 \; , \qquad  \Pi_{ij} = \frac{1}{3c_s^2} \Pi_{00} \; ,
\ee
which are in fact the correct entries of $\langle \partial_\mu \pi \partial_\nu \pi \rangle$ for the case at hand,  as can be easily checked by rescaling the same quantity computed for a free relativistic scalar field by the appropriate powers of $(2P')$ and $c_s$ as dictated by dimensional analysis.

As another simple check, by using the explicit expression for $Z^{\mu\nu}$ \eqref{Zmn}, we can notice that
\be \label{ZPi}
Z^{\mu\nu} \Pi_{\mu\nu} = 0 \; ,
\ee
as it should be:  the linearized equations of motion for $\pi$ read $Z^{\mu\nu} \partial_\mu \partial_\nu \pi = 0 $, and so on any translationally invariant state one should have $Z^{\mu\nu} \langle \partial_\mu \pi \partial_\nu \pi \rangle = 0$ (cf.~\cite{Kourkoulou:2021ksw}).

For what follows, we need the following contractions of $\Pi_{\mu\nu}$:
\begin{align}
[V \Pi]_{\nu}  \equiv V^\mu  \, \Pi_{\mu\nu} & = - \alpha c_s^2 \, X \Big(4 y X  \, u_\nu + \big(X +3(1-c_s^2)y^2 \big) \,V_\nu \Big)\\
[VV\Pi]  \equiv V^\mu V^\nu \, \Pi_{\mu\nu} & =  \alpha c_s^2 X^2 \big(X -(1+ 3c_s^2)y^2  \big)\\
[\Pi]  \equiv \eta^{\mu\nu} \, \Pi_{\mu\nu} & =  \alpha (1-c_s^2) X \big( X - (1+3 c_s^2)  y^2 \big) \; .  \label{[Pi]}
\end{align}

\subsection{The \texorpdfstring{$U(1)$}{} current}
Let's now consider the expectation value of the $U(1)$ current associated with the shift symmetry $\psi \to \psi + {\rm const}$. By applying  Noether's theorem to the superfluid Lagrangian \eqref{eq:superfluid}, we get the current operator
\be \label{J}
J^\mu = \frac{\partial {\cal L}}{\partial (\partial_\mu \psi)} =  - 2 P' \, \partial^\mu \psi \; .
\ee

Expanding in phonon fluctuations around a given background, the scalar takes the form $\psi = \bar \psi + \pi$, with $\partial_\mu \bar \psi \equiv V_\mu$. 
Our free-phonon approximation corresponds to keeping terms up to second order in $\pi$. Recalling that $P$ is a function of $X = - (\partial \psi)^2$, and that $\langle \partial_\mu\pi \rangle = 0$  when evaluated on a translationally invariant state or density matrix, we get
\be
\langle J_\mu \rangle = \bar{J}_{\mu} + J^{(2)}_{\mu}  \; ,
\ee 
where $\bar{J}_{\mu}$  is the background contribution --- the expression \eqref{J} evaluated on the background $\partial _\mu \bar{\psi} = V_\mu$ --- and $J^{(2)}_{\mu}$ is the quadratic correction coming from the thermal gas of phonons:
\begin{align}
J^{(2)}_{\mu} & = \langle \, 4 P'' \, V^\alpha \partial_\alpha \pi \partial_\mu \pi + 2 P'' (\partial \pi)^2 \,  V_\mu  - 4 P''' \,  V^\alpha V^\beta \partial_\alpha \pi  \partial_\beta \pi \, V_\mu \rangle \\
& = 2 \Big( P'' \, \big(2 V^\alpha \delta^{\beta}_{\mu}+  V_\mu \eta^{\alpha\beta} \big) - 2 P''' \,  V^\alpha V^\beta V_\mu  \Big) \, \Pi_{\alpha\beta} \\
& = 2 \Big(P'' \, \big(2 [V\Pi]_\mu + [\Pi] V_\mu \big) - 2 P''' \, [VV \Pi] V_\mu \Big)  \; .
\end{align}

Given the results of the previous subsection, we  get
\be
J^{(2)}_{\mu} = J_u u_\mu + J_V V_\mu \; ,
\ee
with
\begin{align}
J_u & =    \frac{2 \pi^2 T^4}{45}  \frac{c_s (1-c_s^2) \, X^2 y}{\big[X-(1-c_s^2)y^2  \big]^3} \\
 J_V & =  - \frac{\pi^2 T^4}{90}  \, \frac{X}{c_s} \, \frac{ X \frac{d c^2_s}{d X} \big(X- (1+3c_s^2) y^2 \big)
 - 4c_s^2 (1-c_s^2) y^2 }{ \big[X-(1-c_s^2)y^2  \big]^3}
\end{align}
where we have expressed the derivatives of $P$ in terms of the speed of sound and its $X$-derivative, see eq.~\eqref{eq:sound}.

\subsection{The stress-energy tensor}

We can now do the same for stress-energy tensor. From the superfluid Lagrangian \eqref{eq:superfluid}, we get
\begin{equation}\label{eq:Tsup}
    T_{\mu\nu} = 2 P' \,  \partial_\mu \psi \, \partial_\nu \psi + \eta_{\mu\nu} P \; .
\end{equation}
Expanding up to quadratic order in phonon fluctuations, and taking the expectation value on our thermal state, we get
\be 
\expval{T_{\mu\nu}} = \bar{T}_{\mu\nu} + T^{(2)}_{\mu\nu} \; ,
\ee
where $\bar{T}_{\mu\nu}$  is the background contribution --- the expression \eqref{eq:Tsup} evaluated on the background ---
%
and $T^{(2)}_{\mu\nu}$ is the quadratic correction coming from the thermal gas of phonons. Neglecting the expansion of the second term in \eqref{eq:Tsup}, which at quadratic order vanishes thanks to \eqref{ZPi}, we have
\begin{align}
T^{(2)}_{\mu\nu} & = \langle \, 2 P ' \, \partial_\mu \pi \partial_\nu \pi - 8 P'' V^\alpha \,   \partial_\alpha \pi
\partial_{(\mu} \pi  \, V_{\nu)} 
+   \big( 4P''' \,  V^\alpha V^\beta  - 2 P'' \, \eta^{\alpha\beta}\big) \partial_\alpha \pi \partial_\beta \pi \, V_\mu V_\nu \, 
\rangle  \nonumber \\
& = 2 \big( P' \, \Pi_{\mu\nu} - 4 P'' \,  [V \Pi]_{(\mu} V_{\nu)} + \big( 2 P''' \, [VV \Pi] - P'' \, [\Pi] \big) V_\mu V_\nu \big) \\
& = T_u \, u_\mu u_\nu + T_V \, V_\mu V_\nu + T_\eta \, \eta_{\mu\nu} \; , 
\end{align}
with
\begin{align}
T_u & = \frac{2 \pi^2 T^4}{45}  \frac{c_s\, X^3}{\big[X-(1-c_s^2)y^2  \big]^3}  \\
T_V & = \frac{\pi^2 T^4}{90}  \, \frac{X}{c_s} \, \frac{ X \frac{d c^2_s}{d X} \big(X- (1+3c_s^2) y^2 \big)
 - 4c_s^2 (1-c_s^2) y^2 }{ \big[X-(1-c_s^2)y^2  \big]^3} \\
T_\eta & = \frac{ \pi^2 T^4}{90}  \frac{c_s\, X^2}{\big[X-(1-c_s^2)y^2  \big]^2}  \label{Teta}
\; .
\end{align}

Notice that the mixed $u_{(\mu} V_{\nu)}$ tensor structure, which was there  in intermediate steps, completely disappeared from $T_{\mu\nu}$. This is quite nontrivial, and it happens to be a necessary condition for the matching to the EFT to be possible in the first place. We will see why in the next section.
Similarly, despite their complicated structures, $T_V$ and $J_V$ happen to be exactly opposite of each other. Again, were this not the case, it would be impossible to match to the EFT. 

\section{Matching to the two-fluid effective theory}\label{EFT}

We now consider how to encode these thermal corrections into a low-energy effective theory, as described in \cite{Nicolis:2011cs}. The main idea is to describe the thermal phonon gas as a normal fluid, interacting with the underlying superfluid. To lowest order in derivatives, the effective action must take the form \eqref{f}, where $X$, $y$, and $b$ are defined in \eqref{sf blocks}, and the $\phi^I$'s can be interpreted as the comoving coordinates of the phonon gas fluid elements. 

The question then is whether there exists an
\be
F(X,y,b) = P(X) + \delta F(X,y,b) \; ,
\ee 
that describes all the thermal corrections that we derived above. What we mean is this: from the superfluid $P(X)$ at finite temperature we derived the thermal corrections to physical quantities such as the $U(1)$ current and the stress-energy tensor; is there an effective finite-temperature action $F$ that reproduces the same corrections directly from Noether's theorem, without doing any further finite-temperature computations? 

The $U(1)$ current and the stress tensor associated with an $F(X, y, b)$ are  \cite{Nicolis:2011cs}
\begin{align}
J_{\mu} & = F_y u_\mu -2  F_X \partial_\mu \psi \label{J EFT}\\
T_{\mu\nu} & = (F_y y - F_b b) u_{\mu} u_\nu + 2 F_X \partial_\mu \psi \partial_\nu \psi  + (F - F_b b) \eta_{\mu\nu} \; ,
\label{Tmn EFT}
\end{align}
where the subscripts on $F$ denote derivatives with respect to its arguments, and 
\be
u^\mu = \frac{1}{ b} \,  \epsilon^{\mu \alpha \beta \gamma} \, \partial_\alpha \phi^1 \partial_\beta \phi^2 \partial_\gamma \phi^3
\ee
is the four-velocity of the normal fluid (= phonon gas) component. We can already see the importance of the two ``accidents" alluded to at the end of the previous section: there is no mixed $u_{(\mu} \, \partial_{\nu)}\psi$ structure in $T_{\mu\nu}$, and the coefficient of the $\partial_{\mu} \psi$ term in  $J_\mu$ matches that of the $\partial_\mu \psi  \partial_\nu \psi$ term in $T_{\mu\nu}$.

Notice that, up to an overall arbitrary factor, the variable $b$ has to be interpreted as the entropy density $s$
\cite{Dubovsky:2011sj, Nicolis:2015sra}. Let's thus identify the two, and use our results in sect.~\ref{partition} to trade in the temperature for the entropy density $b$:
\be \label{b=s}
b = s = 4 z =  \frac{2 \pi^2 T^3}{45}\frac{c_s \, X^2}{\big[X- (1-c_s^2 ) y^2\big] ^2} \quad \Rightarrow \quad T(X,y,b) 
= \bigg[\frac{45}{2 \pi^2}\Big(1- (1-c_s^2 ) \frac{y^2}{X} \Big)^{2}  \frac{b }{ c_s } \bigg]^{1/3} \; .
\ee

Let's further assume that, by dimensional analysis, at low temperatures the thermal correction $\delta F$ to the effective Lagrangian scales as $T^4$,
\be
\delta F (X,y,b) = \chi(y,X) \, {T^4(X,y,b)} \; ,
\ee
where $\chi$ is a dimensionless function of $X$ and $y$ only, not of $b$, the reason being that it is the entropy density $b$ that is playing the role of the thermal control parameter: low temperature means low $b$, $b \sim T^3$.

Then, by matching one of the entries of our observables, say $T_\eta $ in \eqref{Teta} with the corresponding one in \eqref{Tmn  EFT}, we can immediately determine $\chi(y,X)$:
\be
\chi(y,X)  = -\frac{\pi^2}{30} \, c_s(X)  \, \bigg[1- \big(1-c_s^2(X) \big)\frac{y^2}{X}  \bigg]^{-2} \; ,  
\ee
where we are emphasizing that $c_s$ is, in general, a function of $X$.
We can then check whether such a choice of $\chi(y,X)$ works for all other entries of $J_\mu$ and $T_{\mu\nu}$. Amazingly enough, it does. This is of course a highly nontrivial check of our results and of the general framework. 

In conclusion, all low-temperature thermal corrections to physical observables for a superfluid can be encoded in a two-fluid effective field theory with effective Lagrangian
\be
{\cal L}  = F(X,y,b) = P(X)  - \tilde 1 \, \bigg[ \frac{b^4 }{ c_s(X) } \Big(1- \big(1-c^2_s(X) \big) \frac{y^2}{X} \Big)^{2} \bigg]^{1/3} \; ,
\ee
where $\tilde 1$ is a number that happens to be very close to one \cite{Carter:1995if},
\be
\tilde 1 \equiv \bigg( \frac{1215}{128 \pi^2} \bigg)^{1/3} \simeq 0.99 \; .
\ee
Up to a different normalization for $b$, this is exactly the same result as derived in \cite{Nicolis:2011cs}---see eq.~\eqref{eq:finT} and coincides with (8.4) of \cite{Carter:1995if}.

\newpage
\section{General dimensions}
Despite the tediousness of our computations, it is not difficult to generalize them for a generic number of spatial dimensions $d \neq 3 $.
The spatial dimensionality enters:
\begin{itemize}
\item
The integral defining the partition function, eq.~\eqref{eq:logZderiv}. For general $d$, this can be done by first expanding the log, and then by resumming the series. The partition function then becomes
\begin{equation}\label{z general d}
    z(T, V_\mu) \to c(d) \, \frac{(Z^{00})^d}{\beta^d \, \sqrt{\det G} \big(1-| \vec C|^2 \big)^{\frac{d+1}2}} = 
c(d) \, T^d \, \frac{\sqrt{-\det A}}{ \big( A_{00} \big)^{\frac{d+1}2}}  \; ,
\end{equation}
with 
\be
c(d) \equiv  \frac{ \Gamma(d) \zeta(d+1)}{2^{d-1} \, \pi^{\frac{d}{2}} \Gamma(\frac{d}{2})}  \; .
\ee
This eventually leads, in particular, to modifying our $\Pi_{\mu\nu}$ in \eqref{Pi result} as
\be
\Pi_{\mu\nu} \to -c(d) \cdot T^d \cdot \sqrt{-\det A} \cdot \frac{\big(A_{\mu\nu} A_{\alpha \beta} - (d+1) A_{\alpha \mu} A _{\beta\nu} \big) u^\alpha u^\beta }{ \big( A_{\gamma \delta} \, u^\gamma u^\delta \big)^3 } \; 
\ee
and our entropy density in \eqref{s} as
\be
s \to (d+1) z(T, V_\mu) \; .
\ee

\item
The trace of $\eta_{\mu\nu}$, for instance in \eqref{[Pi]}, which becomes
\be
\eta_{\mu\nu} \eta^{\mu\nu} \to d+1 \; .
\ee
\end{itemize}

Instead of modifying all of our formulae to accommodate for a generic $d$, we report here only the effective theory's Lagrangian, from which the current and the stress tensor can be straightforwardly derived through \eqref{J EFT} and \eqref{Tmn EFT}:
\be \label{EFT general d}
{\cal L} = P(X) - \tilde c(d) \,  \frac{b^{\frac{d+1}{d}} }{ c^{\frac{1}{d}}_s(X) } \Big(1- \big(1-c^2_s(X) \big) \frac{y^2}{X} \Big)^{\frac{d+1}{2d}}  \; ,
\ee
where
\be
\tilde c(d) \equiv \frac{d}{(d+1)^{\frac{d+1}d} c(d)^\frac{1}{d}} \; ,
\ee
and we have kept the same normalization as before for the $b$ variable, $b = s$.

\section{Special cases}

We now consider a number of physically relevant limits. These will also serve as checks of our results.

\subsection{Fluids at relative rest}
A  simple case to consider is that in which the underlying superfluid and the phonon gas are at rest with respect to each other,
\be
u^\mu =(1, \vec{0}) \; , \qquad V_\mu = \mu \,   (1, \vec{0}) = - \mu \, u_\mu \; ,
\ee
where $\mu$ is the chemical potential. Then, we have $X = \mu^2$ and $y = \mu$, and the current and the stress-energy tensor simply become
\begin{align}
\langle J_\mu  \rangle & =\bar{J} _\mu  + \frac{\pi^2}{30} \, \frac{T^4}{\mu c_s^3} \, \frac{d \log c_s^2}{d \log \mu^2}\,  u_\mu \\
\langle T_{\mu\nu}  \rangle&  = \bar{T} _{\mu\nu} +  \frac{\pi^2}{90} \, \frac{T^4}{c_s^3} 
\bigg[\Big( 4- 3 \frac{d \log c_s^2}{d \log \mu^2} \, \Big)u_\mu u _\nu +  \eta_{\mu\nu} \bigg]
\end{align}
These correspond to the following thermal contributions to the number density, energy density, and pressure:
\be
\Delta n =  -\frac{\pi^2}{30} \, \frac{T^4}{\mu c_s^3} \, \frac{d \log c_s^2}{d \log \mu^2} \; ,\qquad
\Delta \rho = \frac{\pi^2}{30} \, \frac{T^4}{c_s^3} \Big( 1- \frac{d \log c_s^2}{d \log \mu^2} \, \Big) \; , \qquad
\Delta p =  \frac{\pi^2}{90} \, \frac{T^4}{c_s^3} \; ,
\ee
which correct the background values
\be \label{n0 etc}
\bar{n} = 2 P'(\mu^2) \mu \; , \qquad \bar{\rho} = 2 P'(\mu^2)  \mu^2 - P(\mu^2) \; , \qquad \bar{p} = P(\mu^2) \; .
\ee

Given that in this case the entropy density \eqref{b=s} is simply
\be \label{easy s}
s = \frac{2 \pi^2 T^3}{45 \, c_s^3} \; ,
\ee
it is straightforward to verify that the standard thermodynamic relationships hold:
\be \label{thermo relations}
\rho + p = \mu \, n + T \, s \; ,  \qquad d p = s \, dT  +  n \, d \mu  \; .
\ee

\newpage
\subsection{Non-relativistic superfluids and the mass of sound}
It is interesting to consider the non-relativistic limit.  To this end, it is useful to reinstate the speed of light, $c$, and expand the EFT action and our observables in inverse powers of $c$, keeping terms of order $c^2 $ ($\sim$ mass density) and of order one ($\sim$ energy density.) From the field theory side, this has been carried out systematically in \cite{Nicolis:2017eqo, Esposito:2018sdc}. For our purposes here, it is more interesting to do this directly at the level of the thermal corrections we have computed for $J_\mu$ and $T_{\mu\nu}$.

Considering for simplicity only the case of the last subsection---i.e., no relative motion between the superfluid and the normal fluid---the replacements we have to perform are quite simple \cite{Nicolis:2017eqo}:
\begin{enumerate}
\item Denote by $m$ the mass associated with one unit of charge as counted by the number density $n$. For instance, for liquid helium-4, if $n$ counts the number of helium atoms per unit volume,  $m$ is the mass of one helium atom.
\item Replace the relativistic chemical potential by $\mu = mc^2 + \mu_{\rm NR}$, where $\mu_{\rm NR}$ is the usual chemical potential of non-relativistic statistical mechanics.
\item From now on, speeds are not dimensionless, and $c_s \ll c$. The chemical potential $\mu_{\rm NR}$ has units of energy, and is typically of order $m c_s^2$.  Our low-temperature limit corresponds to $T \ll \mu_{\rm NR}$.
$T$ has units of energy, and $\rho$ and $p$ of energy density.
\item For all observables, expand in powers of $1/c^2$, and stop at zeroth order. 
\end{enumerate}
If we carry this out and  use the zero-temperature relations \eqref{n0 etc}, we find that the thermal corrections of the last section reduce to
\be
\Delta n \simeq  -\frac{\pi^2}{30} \, \frac{T^4}{m c_s^5} \, \frac{d \log c_s}{d  \log\rho_m} \; ,\qquad
\Delta \rho \simeq \frac{\pi^2}{30} \, \frac{T^4}{c_s^3} \Big( 1- \frac{c^2}{c_s^2} \frac{d \log c_s}{d  \log \rho_m} \, \Big) \; , \qquad
\Delta p =  \frac{\pi^2}{90} \, \frac{T^4}{c_s^3} \; ,
\ee
where $\rho_m$ is the superfluid's mass density, $\rho_m = m \, \bar{n}$.

The surprising feature of these formulae is the term  of order $c^2$ in $\Delta \rho$: it must be interpreted as a thermal correction to the {\it mass} density. 
To understand its origin, recall that free phonons obey the Bose-Einstein distribution,
\be
N_\pi(\vec p) = \frac{1}{e^{E_\pi(\vec p)/T} - 1} \; , \qquad E_\pi (\vec p)\equiv  c_s |\vec p|  \; ,
\ee
and so $\Delta n$ and $\Delta \rho$ above can be written as phase-space integrals as
\begin{align}
\Delta n & = \int \frac{d^3 p}{(2\pi)^3} \, N_{\pi}(\vec p) \, \frac{m_\pi (\vec p)}{m} \\
\Delta \rho & = \int \frac{d^3 p}{(2\pi)^3} \, N_{\pi}(\vec p) \, \big[ \,  m_\pi(\vec p) c^2 + E_\pi(\vec p) \, \big] \; ,
\end{align}
with
\be
m_\pi (\vec p)\equiv - \frac{d \log c_s}{d  \log \rho_m}  \frac{E_\pi (\vec p)}{c_s^2} \; .
\ee
All this is equivalent to saying that each phonon carries with it a mass $m_\pi (\vec p)$, and so a fraction $m_\pi(\vec p)/m$ of charge.

This is in perfect agreement with the results of \cite{Nicolis:2017eqo, Esposito:2018sdc}, where the mass carried by classical sound wave packets and quantum phonons was computed by other means.
On the other hand, the comparison with the classic results by Landau---see e.g.~\cite{Khalatnikov:106134}---is more subtle. We will explore it elsewhere. 

\subsection{Cherenkov phonons and instabilities}\label{Cherenkov}

Landau's original criterion for the breakdown of superfluidity---which turned out to be too weak because of its neglecting the rotons (for liquid helium-4) and vortex rings (for a general superfluid)---is essentially based on the phenomenon of Cherenkov emission: at high enough relative speeds between the superfluid and a container, it becomes energetically possible to excite phonons, and so the superfluid will lose momentum to such phonons, thus effectively developing a viscosity.

A similar phenomenon must be possible in our case. If we do thermodynamics in the lab frame, we are implicitly saying that we have an apparatus at rest in the lab frame that can exchange energy with the superfluid, so as to keep the temperature constant. If the superfluid moves faster than $c_s$ relative to  such an apparatus, Cherenkov emission of phonons becomes energetically possible. If one tried to keep the temperature and relative velocity constant in such a situation, one would find it impossible to reach thermal equilibrium: there would be a constant rate  of phonon production per unit volume.

As explained in \cite{Dubovsky:2005xd}, in the rest frame of the apparatus the possibility of Cherenkov emission must show up as a violation of the stronger stability criterion \eqref{stronger stability}. Let us check that this is indeed the case.
The superfluid's four velocity, for time-like $V^\mu$, is
\be
u^\mu_{\rm sf} = -\frac{V^\mu}{\sqrt{X}} \; ,
\ee
where the minus upfront is conventional. We can thus write $V^\mu$ as
\be
V^{\mu} = -\sqrt{X} \gamma(\vec v) (1, \vec v) \; ,
\ee
where $\vec v$ is the physical velocity of the superfluid in the lab frame, which, without loss of generality, we are taking to have $u^{\mu}=(1,\vec 0)$.

Plugging this $V^\mu$ into \eqref{Zmn} we find
\be
Z^{00} \propto c_s^2 + \gamma^2(\vec v) \, (1-c_s^2) \; , \qquad Z^{ij} \propto \gamma^2(\vec v) (1-c_s^2) v^i v^j - c_s^2 \,\delta^{ij} \; ,
\ee
where we dropped a common positive prefactor. While $Z^{00}$ cannot become negative as long as speeds ($v$ as well as $c_s$) are subluminal, $-Z^{ij}$ can develop a negative eigenvalue for large enough $v$,
\be
-Z^{ij} \, \hat v^i \hat v^j \propto \gamma^2(\vec v) \, (c_s^2-v^2) \; ,
\ee
thus violating \eqref{stronger stability} precisely when $v$ exceeds the speed of sound! In conclusion, there can be an instability at large relative velocities, and it is related to the possibility of emitting Cherenkov phonons.

How does all this show up in our finite-temperature results? Our computation of the partition function assumes that the system is stable, otherwise the  integral that defines the partition function does not converge. So, there is no direct sign of an instability in our final results. However, we can look at what happens when we {\it approach} the unstable region: it should become easier and easier to excite thermal phonons, resulting in a divergent specific heat as $v$ approaches $c_s$ from below.

The easiest way to see this is to look at our result for the entropy density, eq.~\eqref{b=s}. Using the same variables as above, we have
\be
s =  \frac{2 \pi^2 T^3}{45}\frac{c_s}{\gamma^4(\vec v) \, (c_s^2-v^2) ^2} \; ,
\ee 
which, indeed, diverges for $v$ approaching the speed of sound.

\subsection{The purely spacelike case}
One of the motivations of this work was to have a derivation of the thermodynamic properties of a superfluid-like system that is valid even when the four-velocity of the superfluid formally becomes spacelike. The reason we are interested in such an exotic system will be examined elsewhere \cite{kourkoulou}. For the time being, we just display what our thermal corrections look like in this case. 

For simplicity, let's consider a ``purely spacelike" superfluid, but still with a time-like lab frame, that is
\be
V_\mu = (0, \vec V) \; , \qquad u^{\mu} = (1, {\vec 0}) \; .
\ee
In this case we have $X = - V^2$ and $y=0$, and so for our thermal corrections we get
\be
J^{(2)}_0 = 0 \; , \qquad \vec J^{(2)} = - \frac{\pi^2 T^4}{90 \, c_s} \, \frac{d c^2_s}{d X}  \, \vec V
\ee
and
\be
T^{(2)}_{00} = \frac{\pi^2 T^4 \, c_s}{30}  \; ,\qquad
 T^{(2)}_{0i}  = 0 \; , \qquad
 T^{(2)}_{ij}   = \frac{\pi^2 T^4}{90 \, c_s} \, \bigg( \frac{d c^2_s}{d X}  \,  V_i V_j + c_s^2 \delta_{ij} \bigg) \; . 
\ee
\subsection{Conformal superfluids}
Finally, it is immediate to specialize our general results to conformal superfluids, which recently have been a subject of intense study by the `large charge' community \cite{Hellerman:2015nra, Monin:2016jmo, Cuomo:2022kio} and others \cite{Creminelli:2022onn}. Conformal invariance in $d+1$ spacetime dimensions forces $P(X)$ and $c_s(X)$ to take the form
\be
P(X) = c_0 \, X^{\frac{d+1}{2}} \; , \qquad c_s(X) = \frac{1}{d} = {\rm const.} \; ,
\ee
where $c_0$ is a constant.

So, as far as our thermal corrections are concerned, it is enough to use in them this particular value of the sound speed. For example, the $d$-dimensional EFT Lagrangian \eqref{EFT general d} reduces to
\be
{\cal L}_{\rm CFT} = c_0 \, X^{\frac{d+1}{2}}  - \tilde c(d) \, d^{\frac{1}{d}} \,  b^{\frac{d+1}{d}} \Big(1- \big(1-1/d^2 \big) \frac{y^2}{X}\Big)^{\frac{d+1}{2d}} \; ,
\ee
One can easily check that the resulting stress-energy tensor, eq.~\eqref{Tmn EFT}, is traceless, as it should.

Notice that, for large-charge applications, one usually considers  conformal superfluids on a $d$-dimensional sphere. Calling $R$ the radius of such a sphere, and assuming that $\mu = \sqrt X \gg 1/R$ so that the superfluid EFT has a nontrivial regime of validity, our thermal corrections can be trusted only at intermediate temperatures,
\be
1/R \ll T \ll \mu \; . 
\ee
The first inequality ensures that the effects of curvature are negligible---our computations only apply to flat space. The second inequality ensures that thermal phonons have negligible interactions, which we have assumed for our partition function computations.

\section{Discussion}

We have computed the low-temperature thermodynamic properties of a generic superfluid. In particular, we have derived the partition function, the conserved current and stress-energy tensor, and the  low-energy effective action for the corresponding two-fluid model. Our analysis is fully relativistic and model independent. The only input needed is the zero-temperature equation of state, relating, for instance, the pressure to the chemical potential.

We plan to use our results for two applications. The first is the study of certain vortex solutions in a specific UV completion of a superfluid, with an associated puzzle regarding causality when the superfluid's velocity formally becomes superluminal \cite{kourkoulou}. The second is a critical reexamination of Landau's classic derivation of the low-temperature thermal properties of liquid helium \cite{Landau:1941vsj, Landau:1947mij}: in Landau's results,  the mass density associated with the phonon gas --- the density of the `normal' component --- is not quite consistent with the expression for the mass of sound derived in \cite{Nicolis:2017eqo, Esposito:2018sdc}, while our results are in perfect agreement with it. We would like to understand the origin of this discrepancy. Somewhat related to this, it would be interesting to understand how our results would be modified by rotons: for liquid helium, these start dominating thermodynamic quantities at temperatures of order of the transition temperature, but still well below it \cite{Khalatnikov:106134}, that is, to some extent still in the ``low-temperature" regime. Much of our formalism can still be applied, but other things will have to be modified.  

We hope our results will also find useful applications in the case of conformal superfluids, which have recently received considerable attention \cite{Hellerman:2015nra, Monin:2016jmo, Cuomo:2022kio, Creminelli:2022onn}.

\subsection*{Acknowledgements}
We thank Michael Landry for discussions and collaboration on a related project \cite{kourkoulou}.
Our work has been  supported by the US DOE (award number DE-SC011941) and by the Simons Foundation (award number 658906).

\newpage
\bibliographystyle{JHEP.bst}
\bibliography{deriv}

\end{document}